\documentclass[modern]{aastex701}

\shorttitle{Fast binary-lens $\Delta\chi^2$ maps}
\shortauthors{Hippke}

\begin{document}

\title{\texttt{lenscape}: Fast, High-resolution $\Delta\chi^2$ Maps of Microlensing Events}

\author{Michael Hippke}
\affiliation{Sonneberg Observatory, Sternwartestr. 32, 96515 Sonneberg, Germany}
\email{michael@hippke.org}

\begin{abstract}
Planetary microlensing analyses rely on $\Delta\chi^2$ maps across the binary-lens separation $s$ and mass ratio $q$ to locate degenerate solutions and bound parameter uncertainties. Because every grid point requires a full light-curve optimization with finite-source magnifications, constructing these maps is often the primary computational bottleneck in event modeling. Here I present \texttt{lenscape} (\url{https://github.com/hippke/lenscape}), an open-source Python package that accelerates $\Delta\chi^2$ mapping and suppresses local-minimum artifacts through multi-stage neighbor propagation and adaptive contour refinement. Applied to the planetary event MOA-2022-BLG-091, \texttt{lenscape} reproduces the published $50\times50$ map in 14 minutes and generates a $400\times400$ map in under two hours on a desktop computer.
\end{abstract}


\section{Introduction}

Planetary microlensing light curves frequently admit multiple degenerate binary-lens interpretations. To chart this multimodal likelihood surface, modelers evaluate a grid in $(\log s, \log q)$, optimizing the remaining parameters---peak time $t_0$, impact parameter $u_0$, Einstein timescale $t_{\rm E}$, trajectory angle $\alpha$, and normalized source radius $\rho$---at every lattice point. The resulting $\Delta\chi^2$ maps govern both the number of reported solutions and the width of the inferred mass-ratio constraints. Yet their computational cost is substantial. For a 533-cell grid with eight initial trajectory angles per cell, \citet{Shang2025} report ${\sim}4.5$~CPU-hours per light curve when magnifications are interpolated from precomputed maps, and ${\sim}400$~CPU-hours without them. Planetary sensitivity studies require ${\sim}10^4$ such maps per event, totaling ${\sim}4.5\times10^4$~CPU-hours per event and ${\sim}10^8$~CPU-hours for a statistical sample \citep{Shang2025}. Consequently, standard analyses are restricted to coarse grids with independent point-by-point fits that are prone to trapping in local minima. Recent work reduces either the magnification cost, through interpolated precomputed magnification maps \citep{Shang2025, AdaMGrid}, or the grid geometry, through the $(k, h)$ parameterization of \citet{Hall2024}, which concentrates grid cells where central and resonant caustic degeneracies occur.

\begin{figure}[ht!]
\includegraphics[width=\textwidth]{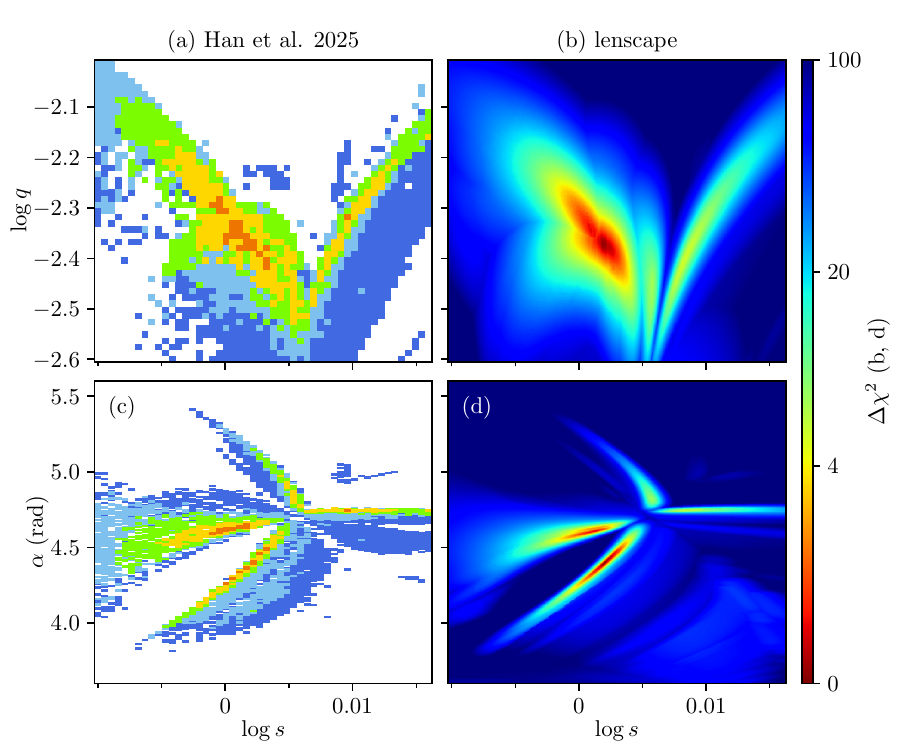}
\caption{Validation on the planetary microlensing event MOA-2022-BLG-091. Panels (a) and (c) show the published $50\times50$ maps \citep[Fig.~2 of][]{Han2025} with discrete color bands at $\Delta\chi^2 < 4, 8, 12, 16,$ and $20$ (red to blue). Panels (b) and (d) display the continuous $\Delta\chi^2$ surfaces computed with \texttt{lenscape} at $400\times400$ resolution. Top row (a, b): $(\log s, \log q)$ plane. Bottom row (c, d): trajectory angle $\alpha$ versus $\log s$, with $\chi^2$ profiled over all remaining parameters (including $q$) at each grid point.}
\label{fig:map}
\end{figure}

\section{Method}

\texttt{lenscape} is implemented in pure Python and just-in-time compiled to machine code via \texttt{Numba} \citep{Lam2015}.

\textit{Magnifications.} Point-source magnifications follow the polynomial root solver of \texttt{VBMicrolensing}\footnote{\url{https://github.com/valboz/VBMicrolensing}, the successor of \texttt{VBBinaryLensing}.} \citep{Bozza2010, Bozza2018, SkowronGould2012}. Where finite-source effects become significant, quadrupole and hexadecapole multipole approximations \citep{Gould2008, Pejcha2009, Cassan2017} are evaluated along images tracked continuously from the point-source roots; full Green's-theorem contour integration is reserved strictly for sources straddling or enclosed by a caustic.

\textit{Optimization at fixed $(s, q)$.} At each grid point, source and blend fluxes for each observatory dataset are solved analytically via linear least squares, while the nonlinear parameters are optimized in the reparameterized basis $(t_0, t_{\rm eff}, t_{\rm E}, t_\star, \alpha)$, where $t_{\rm eff} \equiv u_0 t_{\rm E}$ and $t_\star \equiv \rho t_{\rm E}$. As established for high-magnification events \citep{Dong2006, Gould2014} and implemented in packages such as \texttt{MulensModel} \citep{Poleski2019}, this transformation rectifies the narrow, curved likelihood valleys induced by parameter correlations. A brief Nelder--Mead simplex search first locates the valley floor, after which Levenberg--Marquardt steps rapidly converge to the minimum, analogous to \texttt{RTModel} \citep{Bozza2024}. During the initial simplex phase, residuals are regularized with a Huber loss \citep{Huber1964} to dull the needle-like $\chi^2$ spikes that arise when a point-like source crosses a fold caustic; final convergence and all reported $\chi^2$ values use the exact likelihood. Because the likelihood is strongly multimodal in $\alpha$, de novo fits sweep a dense grid of initial trajectory angles and polish the most promising candidates to convergence.

\textit{Staged map construction.} Independent optimizations at fixed lattice points frequently stall in secondary minima, yielding speckled maps that obscure true degeneracies. \texttt{lenscape} overcomes this through a four-stage hierarchy:
(i)~A coarse cell grid is optimized with $s$ and $q$ free to vary within each cell's boundaries, ensuring narrow likelihood valleys passing between lattice nodes are captured.
(ii)~Adjacent cells exchange trial solutions, triggering a local refit whenever a single objective evaluation indicates an improvement.
(iii)~The fixed-$(s, q)$ lattice nodes are initialized and optimized from the surrounding cell solutions.
(iv)~Solutions are iteratively propagated across neighboring nodes---seeding both resolved and unresolved source branches, which occupy distinct basins of attraction---until no point improves by $\Delta\chi^2 > 1$. Any residual outlier nodes standing above their neighbors undergo a deep multi-start search.

\textit{Hierarchical densification.} High-resolution maps are constructed progressively from coarser grids. Newly inserted midpoint nodes are initialized by interpolating neighboring solutions and accepted immediately if their fitted $\chi^2$ is locally smooth; otherwise, iterative neighbor exchanges reconcile the discrepancy. Optionally, refinement can be restricted to cells intersected by target $\Delta\chi^2$ contours, resolving degeneracy boundaries without full-grid overhead. This $(s, q)$-plane refinement complements the adaptive source-plane magnification grids of \texttt{AdaMGrid} \citep{AdaMGrid}. Applying the same framework at fixed $(\log s, \alpha)$ with $q$ free yields the profile likelihood maps shown in Figure~\ref{fig:map}(d).

\section{Validation and Performance}

Figure~\ref{fig:map} benchmarks \texttt{lenscape} on MOA-2022-BLG-091, a planetary microlensing event exhibiting four distinct degenerate solutions \citep{Han2025}. The continuous \texttt{lenscape} surfaces recover all four minima, closely match the morphology of the published maps, and reproduce the published best-fit $\chi^2$ values to within $\Delta\chi^2 < 1$.

On a 16-core AMD Ryzen 7950X processor, the initial $50\times50$ lattice required 14~minutes. Three successive dyadic refinements ($100\times100$, $200\times200$, and $400\times400$) required 8, 23, and 74~minutes, respectively, bringing the total wall-clock runtime for the $400\times400$ map in Figure~\ref{fig:map}(b) to 119~minutes. Parallel scaling is near-linear across CPU cores. The $50\times50$ map thus costs ${\sim}3.7$~CPU-hours for 2500 cells, or ${\sim}0.09$~CPU-minutes per cell, compared with ${\sim}0.5$ and ${\sim}45$~CPU-minutes per cell for the interpolated and direct pipelines of \citet{Shang2025}, corresponding to speed-ups of ${\sim}6\times$ and ${\sim}500\times$, respectively. This comparison is approximate, as the number of data points, initial trajectory angles, and treatment of $\rho$ differ between the two analyses.

\section{Conclusion}

By combining compiled multipole-accelerated magnification routines with staged neighbor propagation and adaptive refinement, \texttt{lenscape} makes high-resolution $\Delta\chi^2$ maps a desktop computation, without interpolation errors and with suppressed local-minimum artifacts. I invite the community to test and extend the code across the full diversity of microlensing events through issues and pull requests.

\begin{acknowledgements}
AI (Anthropic Claude Opus 5.5) was used to assist in writing portions of the software.
\end{acknowledgements}

\software{\texttt{Numba} \citep{Lam2015}, \texttt{VBMicrolensing} \citep{Bozza2010, Bozza2018}, \texttt{lenscape} (\url{https://github.com/hippke/lenscape})}

\bibliography{rnaas}{}
\bibliographystyle{aasjournal}

\end{document}